\documentclass[12pt,preprint]{aastex}

\slugcomment{Submitted to Astrophysical Journal}

\shorttitle{Solar flare S abundance}
\shortauthors{Sylwester et al.}

\begin{document}

\title{THE SOLAR FLARE SULPHUR ABUNDANCE FROM RESIK OBSERVATIONS}


\author{J. Sylwester, B. Sylwester }
\affil{Space Research Centre, Polish Academy of Sciences, 51-622, Kopernika~11, Wroc{\l}aw, Poland}
\email{js@cbk.pan.wroc.pl,bs@cbk.pan.wroc.pl}

\author{K. J. H. Phillips}
\affil{Mullard Space Science Laboratory, University College London, Holmbury St Mary, Dorking,
Surrey RH5 6NT, U.K.}
\email{kjhp@mssl.ucl.ac.uk}

\and

\author{V. D. Kuznetsov }
\affil{Institute of Terrestrial Magnetism and Radiowave Propagation (IZMIRAN), Troitsk, Moscow, Russia}
\email{kvd@izmiran.ru}

\begin{abstract}
The RESIK instrument on {\em CORONAS-F} spacecraft observed several sulphur X-ray lines in three of its four channels covering the wavelength range 3.8--6.1~\AA\ during solar flares. The fluxes are analyzed to give the sulphur abundance. Data are chosen for when the instrument parameters were optimized.
The measured fluxes of the \ion{S}{15} $1s^2-1s4p$ ($w4$) line at 4.089~\AA\ gives $A({\rm S}) = 7.16 \pm 0.17$ (abundances on a logarithmic scale with $A({\rm H}) = 12$) which we consider to be the most reliable. Estimates from other lines range from 7.13 to 7.24. The preferred S abundance estimate is very close to recent photospheric abundance estimates and to quiet-Sun solar wind and meteoritic abundances. This implies no fractionation of sulphur by processes tending to enhance the coronal abundance from the photospheric that depend on the first ionization potential (FIP), or that sulphur, though its FIP has an intermediate value of 10.36~eV, acts like a ``high-FIP" element.
\end{abstract}

\keywords{Sun: abundances --- Sun: corona --- Sun: flares --- Sun: X-rays, gamma rays  --- line:
identification}

% Section 1
\section{INTRODUCTION}\label{intro}

In previous works \citep{phi10,syl10a,syl10b,syl10c,syl11}, we have used the many hundreds of solar flare and active region X-ray spectra available from the RESIK (REntgenovsky Spektrometr s Izognutymi Kristalami: \cite{syl05}) crystal spectrometer on the {\em CORONAS-F} spacecraft (launched 2001 July~31) to derive the abundances of K, Ar, and Cl, and to examine the continuum emission which can be measured with RESIK. This was done for 2795 spectra collected during nineteen flares between 2002 August and 2003 February. Sulphur lines are prominent in RESIK flare spectra in three of its four channels. These include the He-like S (\ion{S}{15}) $1s^2\,^1S_0 - 1s4p\,^1P_1$ at 4.088~\AA\ (referred to here as $w4$, in channel~2), the H-like S (\ion{S}{16}) Ly-$\alpha$ doublet forming a single feature at 4.729~\AA\ (channel~3), and the He-like S $1s^2 - 1s2l$ ($l = s$, $p$) triplet of lines at 5.04--5.10~\AA\ (channel~4). In addition, the \ion{S}{15} $w3$ ($1s^2\,^1S_0 - 1s3p\,^1P_1$) line at 4.299~\AA\ was occasionally visible in channel~2 (long-wavelength end) or channel~3 (short-wavelength end) for flares with large offset from the optical axis of RESIK. Since all these lines are intense during flares and for non-flaring active regions, several independent sulphur abundance determinations are in principle possible. As in our previous work on flares, we assume that the emitting plasma is isothermal and that the electron temperature $T_e$ and volume emission measure $N_e^2 V$ ($N_e = $ electron density, $V = $ the emitting volume) are given by values ($T_{GOES}$ and $EM_{GOES}$) derived from the flux ratio of X-ray emission in the two channels of {\em GOES}. This assumption appears to hold for the relatively high-temperature \ion{K}{18}, \ion{Ar}{17}, and \ion{Cl}{17} line emission and continuum emission in flare plasmas, as discussed before \citep{phi10,syl10b,syl10c,syl11}. For lower-temperature non-flaring emission it has been found necessary to use a differential emission measure technique \citep{syl10a}. For the He-like and H-like sulphur lines emitted during flares that are studied in this work, the isothermal assumption again appears to be valid as we shall illustrate later. On 2002 December~24, the instrumental parameters of channels~3 and 4 were optimized, resulting in a reduction of crystal fluorescence to the background. As these adjustments affected the S lines occurring in these channels,  we confined our analysis to flares occurring in 2003 January and February and so to a rather smaller number of spectra than considered in our previous work.

Sulphur is of much interest in discussions of the first ionization potential (FIP) effect as its FIP is 10.36~eV, an intermediate value between elements that are ``low-FIP" (elements mostly in ionized form in the photosphere or chromosphere) and those that are ``high-FIP" (elements mostly in neutral form). According to some discussions \citep{fel92,fel00}, low-FIP elements have abundances that are enhanced by as much as a factor of four in the corona over their photospheric abundances. Some of the earlier attempts to explain such abundance enhancements have been discussed by \cite{hen98}; they generally involved fractionation processes operating on ions but not neutral atoms in the photosphere. None of these models gives enhancement amounts so it is difficult to distinguish their relative merits or demerits. An elaborate model \citep{lam04,lam09,lam12} in which acceleration of ions in the photosphere or chromosphere occurs by a ponderomotive force associated with Alfv\'en waves passing into or down from the corona has been advanced. With particular values of magnetic loop lengths and Alfv\'en wave fluxes, enhancement factors for different elements can be evaluated; these can then be compared with observations such as those from RESIK. With refinements in photospheric abundances in recent years \citep{asp09,caf11} and coronal abundances from flare and active region plasmas from RESIK and other spectrometers, much more detailed observations of enhancement factors and their dependence on FIP, if any, are possible than was hitherto possible.

For the case of S, the earlier photospheric abundance determinations (e.g. \cite{gre98}) gave $A({\rm S}) = 7.33$ (on a logarithmic scale in which $A({\rm H}) = 12$); this is larger than more recent values based on improved observational data and modeling including effects due to solar granules; thus, \cite{asp09} now estimate $A({\rm S}) = 7.12 \pm 0.03$ and \cite{caf11} $A({\rm S}) = 7.16 \pm 0.05$. \cite{fel00} give the coronal abundance (more specifically the abundance for quiet coronal regions with temperature $\leqslant 1.4$~MK) as $A({\rm S}) = 7.33$, i.e. a factor 1.5 to 1.6 higher than the recent photospheric abundances. However, the \cite{fel00} abundance is higher than that obtained from flares observed with the {\em Yohkoh} Bragg Crystal Spectrometer: $A({\rm S}) = 6.90$ \citep{flu99}, a value actually {\em less} than the recent photospheric values. \cite{syl10a} measured a factor-of-3 range of abundances from RESIK observations of non-flaring active regions (6.75--7.25), which may reflect time variations; according to \cite{fel00}, the enhancements are expected to be greater for  older active regions.  For solar wind measurements, \cite{rea99} using {\em Wind} data and \cite{von00} using {\em Ulysses} SWICS observations give abundances relative to O ions; assuming the O abundance from \cite{asp09}, ``quiet time" {\em Wind} measurements give $A({\rm S}) =7.15$, similar to the recent photospheric abundances, while the SWICS results indicate a range for the slow solar wind (7.38--7.44) but a constant value for the fast solar wind (7.40), both enhanced over photospheric values. Meteoritic S abundances are also available: \cite{lod03} gives $A({\rm S}) = 7.19\pm 0.04$ for CI carbonaceous chondrites, which of all meteorites are considered to reflect most closely the solar photospheric composition.

In this work, we discuss RESIK flare spectra from the various S lines available to give abundance values, the precision of which we compare. It updates work based on a preliminary calibration of RESIK by \cite{phi03}.

% Section 2
\section{RESIK FLARE SPECTRA AND ANALYSIS}

The instrumental details of RESIK and the data analysis procedure have been given before \citep{syl05,syl10b}, so only a summary is given here. RESIK consisted of a pair of spectrometers, with two bent crystals in each to diffract the incoming solar X-rays and position-sensitive proportional counters to detect the diffracted emission. The nominal wavelength ranges of the four channels of RESIK were 3.40--3.80~\AA\ (channel~1); 3.83--4.27~\AA\ (2); 4.35--4.86~\AA\ (3); and 5.00--6.05~\AA\ (4).  The crystal for channels 1 and 2 was silicon (diffracting plane Si~111), for channels~3 and 4 quartz ($10{\bar1}0$). Thus, the \ion{S}{15} $w4$ line (4.088~\AA) fell in channel~2, the \ion{S}{16} Ly-$\alpha$ line (4.729~\AA) in channel~3, and the \ion{S}{15} $1s^2 - 1s2l$ lines (5.04--5.10~\AA) in channel~4. The \ion{S}{15} $w3$ line (4.299~\AA) line sometimes occurred at the short-wavelength end of channel~3, sometimes at the long-wavelength end of channel~2, and at other times not apparent at all, this depending on the angular offset of the flare from the optical axis of RESIK. As with all solar Bragg crystal spectrometers, a background is liable to be formed by fluorescence of the crystal material by solar X-rays. As described previously, through careful separation of the solar from the fluorescence photons using pulse height analyzer data, the fluorescence background could be entirely eliminated from channels~1 and 2 and greatly reduced for channels~3 and 4. The instrumental settings for channels 3 and 4 were adjusted for optimal background at the end of 2002. Since the \ion{S}{16} Ly-$\alpha$ line, the \ion{S}{15} $1s^2 - 1s2l$ lines, and occasionally the \ion{S}{15} $w3$ line all fell in channels~3 and 4, we selected for analysis only those spectra collected during flares in 2003. Details of the flares are given in Table~\ref{flare_list}. The total number of spectra from the thirteen flares observed was 1448. Though this is a smaller number of spectra from those used in our previous analyses of potassium and argon lines and continuum emission in channels~1 and 2, there are sufficient data to allow determinations of the sulphur abundance from the lines available. An extensive pre-launch assessment of instrument parameters \citep{syl05} resulted in a $\sim 20$~\% absolute calibration, so the S abundance determinations from the various lines available should have a high precision and enable comparison with photospheric abundances.

% Table 1: List of RESIK flares in 2003 (part of Table in Phillips et al. continuum paper)
\begin{deluxetable}{lccccc}
\tabletypesize{\scriptsize} \tablecaption{F{\sc lares with analyzed} RESIK S{\sc pectra} \label{flare_list}}
\tablewidth{0pt}

\tablehead{\colhead{Date} & \colhead{Time of Flare} &  \colhead{{\it GOES} Importance} & \colhead{Location} & \colhead{Number of} & \colhead{S XV $w3$ line}
\\
& \colhead{Maximum (UT)} & & &\colhead{Spectra} & \colhead{Channel 2 or 3} }

\startdata

2003 January    7 & 23:30 & M4.9 & S11E89 &  113 & 3\\
2003 January    9 & 01:39 & C9.8 & S09W25 &  297 & -\\
2003 January   21 & 02:28 & C8.1 & N14E09 &   69 & 3\\
2003 January   21 & 02:50 & C4.0 & N14E09 &   25 & 3\\
2003 January   21 & 15:26 & M1.9 & S07E90 &  290 & 3\\
2003 February   1 & 09:05 & M1.2 & S05E90 &  133 & 3\\
2003 February   6 & 02:11 & C3.4 & S16E55 &   34 & 3\\
2003 February  14 & 02:12 & C5.4 & N12W88 &   44 & -\\
2003 February  14 & 05:26 & C5.6 & N11W85 &   64 & -\\
2003 February  15 & 08:10 & C4.5 & S10W89 &  299 & 2\\
2003 February  21 & 19:50 & C4.3 & N15E01 &   32 & 3\\
2003 February  22 & 04:50 & B9.6 & N16W02 &   19 & -\\
2003 February  22 & 09:29 & C5.8 & N16W05 &   29 & 3\\

\\

\enddata

\end{deluxetable}

\begin{figure}
\epsscale{.80}
\plotone{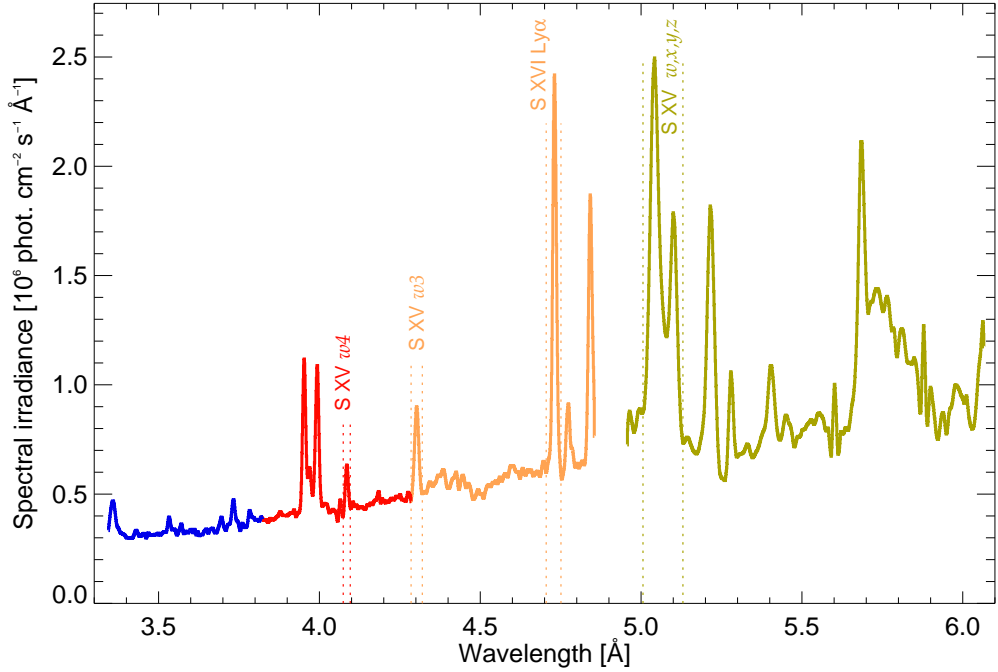}
\caption{Complete RESIK spectrum (channels 1 to 4) taken during the M4.9 flare on 2003 January~7 over the period 23:29--23:32:35 UT. The principal S lines used in the analysis here are indicated, with vertical dashed lines to show the interval over which the line fluxes were estimated. (Color versions of this and other figures in this paper are available in the online journal.)} \label{complete_spectrum}
\end{figure}

Figure~\ref{complete_spectrum} shows a complete spectrum from all four RESIK channels in a 3.5-minute interval during the M4.9 flare on 2003 January~7. Line styles (colors in the on-line journal) indicate the four RESIK channels, and the principal S lines discussed in this work are labeled (most of the remaining lines are due to highly ionized Si and Ar). Values of the wavelength resolution (FWHM) are 8~m\AA\ (channel~1), 9~m\AA\ (2), 12~m\AA\ (3), and 17~m\AA\ (4). The resolution for channel~4 is thus not quite sufficient to resolve the \ion{S}{15} intercombination lines ($x$, $y$) at 5.065~\AA\ from the \ion{S}{15} resonance ($w$) line at 5.039~\AA, unlike some other spectrometers such as {\em Yohkoh} BCS \citep{har96,wat95}, but the summed flux of these lines and the forbidden line ($z$, 5.102~\AA) can still be determined. Fluxes of all the S lines or line groups to be analyzed were estimated from the total emission in the spectral intervals indicated by vertical dashed lines in Figure~\ref{complete_spectrum} and subtracting a portion of continuum (channels~1 and 2) or background (continuum and crystal fluorescence in channels~3 and 4) in neighboring intervals. Thus the flux in the \ion{S}{15} $w4$ line at 4.088~\AA\ appearing in channel~2 was estimated by taking the emission in the 4.075--4.095~\AA\ interval and subtracting the continuum in the 4.064--4.068~\AA\ and 4.112--4.127~\AA\ intervals. An automated routine enabled these measurements to be taken.

% Fig. 1: Complete RESIK channels 1-4 spectra, showing S lines

\begin{figure}
\epsscale{.80}
\plotone{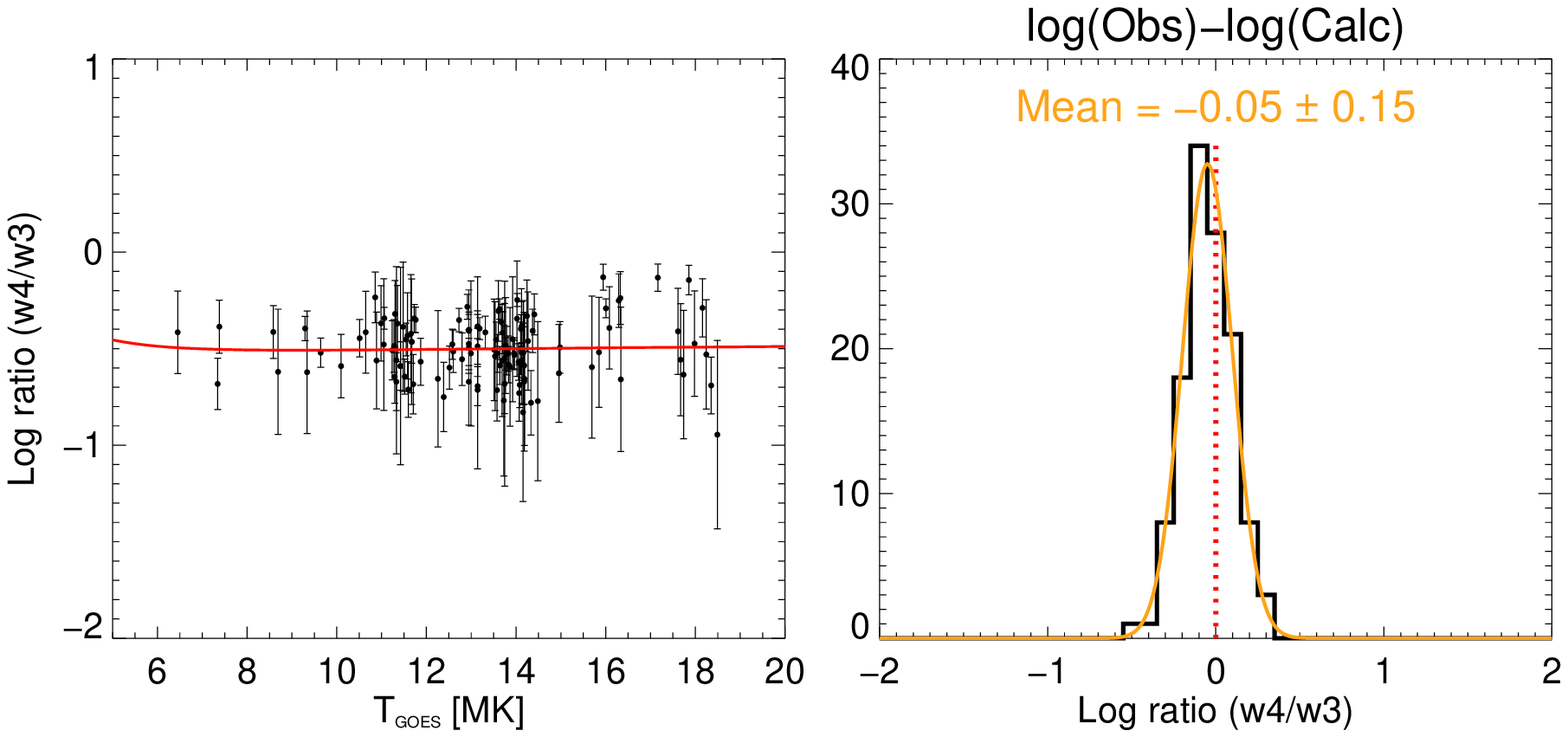}
\caption{Left panel: Logarithm of the \ion{S}{15} $w4$ (channel 2) to $w3$ (channel 3) line flux ratio (error bars indicate statistical uncertainties) as a function of $T_{GOES}$ for 725 spectra collected during eight flares for which the $w3$ line fell at the short-wavelength end of channel~3. The solid curve is the theoretical ratio from {\sc chianti}. Right panel: Number distribution of the logarithm of the ratios. The logarithm of the mean ratio divided by the ratio from {\sc chianti} is indicated with uncertainty.} \label{SXV_w3_ch3_w4_ch2}
\end{figure}

Fluxes of lines falling near channel edges may be subject to uncertainties through anomalous values of crystal reflectivity if crystal curvature effects are significant. This is relevant to the \ion{S}{15} $w3$ line appearing in either channel~2 or 3. By measuring the ratio of the \ion{S}{15} $w3$ and $w4$ line in channel~2 as a function of $T_{GOES}$, this effect can be checked, as can the estimates made of crystal fluorescence relative to solar continuum. Figure~\ref{SXV_w3_ch3_w4_ch2} (left panel) shows this plot for the 725 spectra collected during eight flares for which the $w3$ line fell in channel~3. Figure~\ref{SXV_w3_ch3_w4_ch2} (right panel) shows the number distribution of the ratio relative to that given by the {\sc chianti} database and software package (version 6.0.1: \cite{der97,der09}) on a logarithmic scale; as can be seen, the ratio is equal to that given by {\sc chianti} to within uncertainties. We conclude that measurements of the $w3$ flux when this line falls in channel~3 are available for sulphur abundance estimates, and that any reflectivity anomalies arising from edge effects can be neglected.

% Fig. 2: S XV w4 (channel 2) to w3 (channel3) ratio

% Section 3
\section{ABUNDANCE OF S}\label{S_abundance}

Other than limiting the observational data to spectra after the adjustment of RESIK's instrument parameters in early 2003, our procedure for estimating the sulphur abundance is as described in our earlier work for K, Ar, and Cl abundances from RESIK solar flare spectra \citep{syl10b,syl10c,syl11}. In summary, this is to assume that the emitting plasma is isothermal with temperature and volume emission measure given by $T_{GOES}$ and $EM_{GOES}$. The estimated fluxes of each of the S lines to be analyzed divided by $EM_{GOES}$ are plotted against $T_{GOES}$. If the emitting plasma is isothermal, the points should cluster around the theoretical contribution or $G(T_e)$ function for the line, including any unresolved line components such as dielectronic satellite lines, where $G(T_e)$ is given by \citep{phi08}

% Eq. 1
\begin{equation}
G(T_e) =  \frac{N({\rm S}^{+n}_i)}{N({\rm S}^{+n})} \frac{N({\rm S}^{+n})}{N({\rm S})} \frac{N({\rm S})}{N({\rm H})} \frac{N({\rm H})}{N_e} \frac{A_{i0}}{N_e} \,\,\,\,\,\,{\rm cm}^3\,\,{\rm s}^{-1}
\end{equation}

\noindent where $N({\rm S}^{+n}_i)$ is the population of the excited ($i$th) level of the ion S$^{+n}_i$ ($n=14$ or 15), $N({\rm S}^{+n})/{N({\rm S})}$ is the ion fraction, taken in our case from \cite{bry09}. The abundance of S relative to H, $N({\rm S})/N({\rm H})$, is chosen to be both the photospheric value given by \cite{asp09}, viz. $N({\rm S})/N({\rm H}) = 1.3 \times 10^{-5}$ (equivalent to $A({\rm S}) = 7.12$) and the coronal value given by \cite{fel00}, $N({\rm S})/N({\rm H}) = 2.14 \times 10^{-5}$ ($A({\rm S}) = 7.33$). Other quantities in Equation~(1) are $N_e$ (electron density), and $A_{i0}$ (transition probability from level $i$ to the ground state), and $N({\rm H})/N_e$ which is taken to be 0.83. The excitation rates for both H-like and He-like S, defining the upper-level populations, are derived from $R$-matrix calculations taking account of auto-ionizing resonances \citep{kim00,agg91}. The small contribution to $G(T_e)$ made by unresolved dielectronic satellite lines was taken from {\sc chianti} (version~6.0.1) data; this decreases with $T_e$. For each S line, the S abundance and uncertainty are determined from the peak and width of the distribution of individual abundance estimates. In the following, we give the plots against $T_{GOES}$ of the observed line fluxes divided by $EM_{GOES}$ and the S abundance determinations as histogram plots for each of the lines analyzed.

\begin{figure}
\epsscale{.80}
\plotone{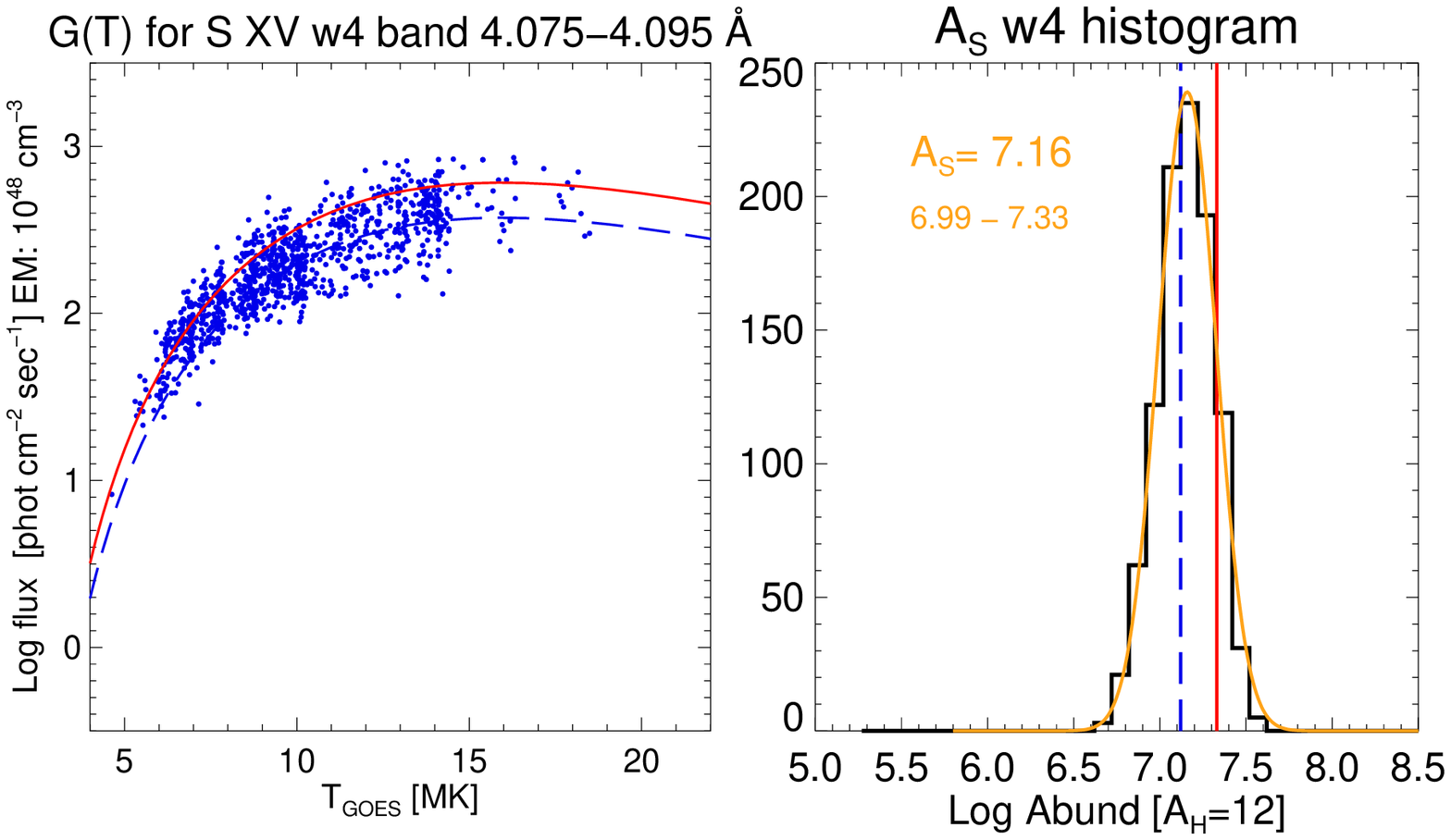}
\caption{Left panel: Measured \ion{S}{15} $w4$ line emission (in RESIK channel~2 range 4.075--4.095~\AA) divided by EM$_{GOES}$ plotted against $T_{\rm GOES}$. In this and all following figures, the dashed blue curve is the theoretical $G(T)$ function for the photospheric S abundance (\cite{asp09}: $A({\rm S}) = 7.12$), the red curve for a nominal coronal abundance (\cite{fel00}: $A({\rm S}) = 7.33$). Points from 1448 RESIK spectra during thirteen flares in 2003 are included. Right panel: Distribution of abundance determinations with best-fit Gaussian curve. From the peak and width of the Gaussian, the S abundance is determined to be $A({\rm S}) = 7.16 \pm 0.14$. The vertical blue dashed and red lines correspond to the photospheric and coronal abundance respectively. } \label{SXV_w4_GofT_ch2}
\end{figure}

Of the sulphur lines available in RESIK spectra for sulphur abundance estimates, the \ion{S}{15} $w4$ line at 4.088~\AA\ is likely to be the most favorable since it falls in channel~2, one of the best characterized channels of RESIK. The plots relevant to this line are given in Figure~\ref{SXV_w4_GofT_ch2}, where in the left panel line fluxes divided by $EM_{GOES}$ in units of $10^{48}$~cm$^{-3}$ are plotted for all 1448 spectra in the thirteen flares of Table~\ref{flare_list} and the distribution of S abundance determinations from these points are given (right panel). The legend of this plot gives the peak and width (FWHM) from the best-fit Gaussian curve; from these, the sulphur abundance is determined to be $A({\rm S}) = 7.16 \pm 0.17$.

% Fig. 3: S XV w4 line, channel 2

\begin{figure}
\epsscale{.80}
\plotone{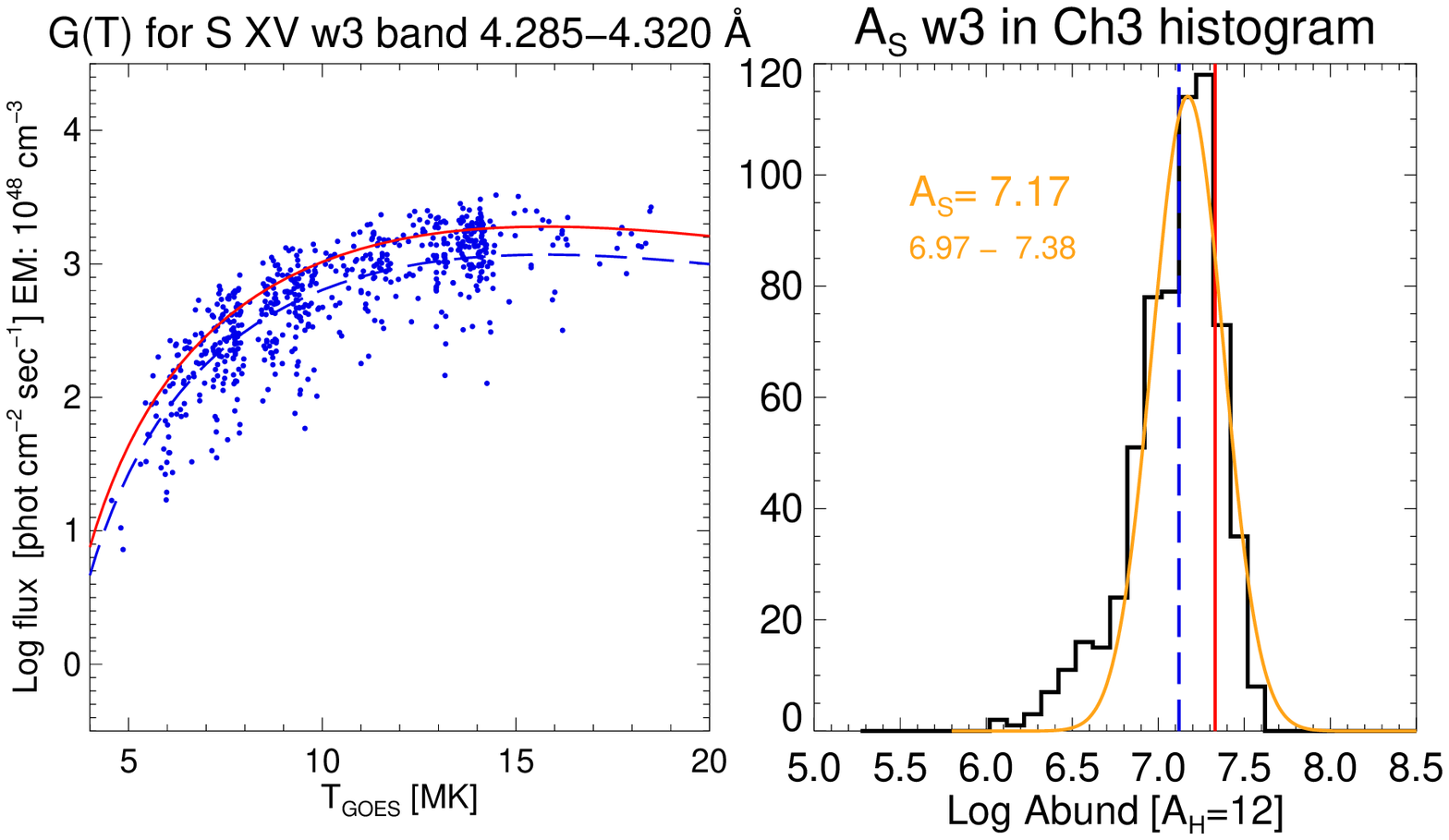}
\caption{ Left panel: \ion{S}{15} $w3$ line emission (line emission in range 4.285--4.320~\AA) plotted against $T_{\rm GOES}$. The points are based on analysis of 554 spectra during the eight flares for which there are $> 20$ photon counts in the $w3$ line which fell at the short-wavelength end of channel~3. Right panel: Histogram of abundance determinations with best-fit Gaussian curve. Here the value of $A({\rm S}) = 7.17 \pm 0.17$. } \label{SXV_w3_GofT_ch3}
\end{figure}

Sulphur abundance estimates can also be made from the \ion{S}{15} $w3$ line when, for 725 spectra in eight flares, this line falls at the short-wavelength end of channel~3. As Figure~\ref{SXV_w3_ch3_w4_ch2} indicates, the $w3$ to $w4$ line flux ratio is nearly constant, following the theoretical ratio as calculated from {\sc chianti}. The estimated S abundance should therefore be similar to that obtained from the \ion{S}{15} $w4$ line. This is indeed the case, as is indicated by Figure~\ref{SXV_w3_GofT_ch3}. Here the distribution of abundance estimates indicates that $A({\rm S}) = 7.17 \pm 0.20$.

% Fig. 4 : S XV w3 line in channel 3

\begin{figure}
\epsscale{.80}
\plotone{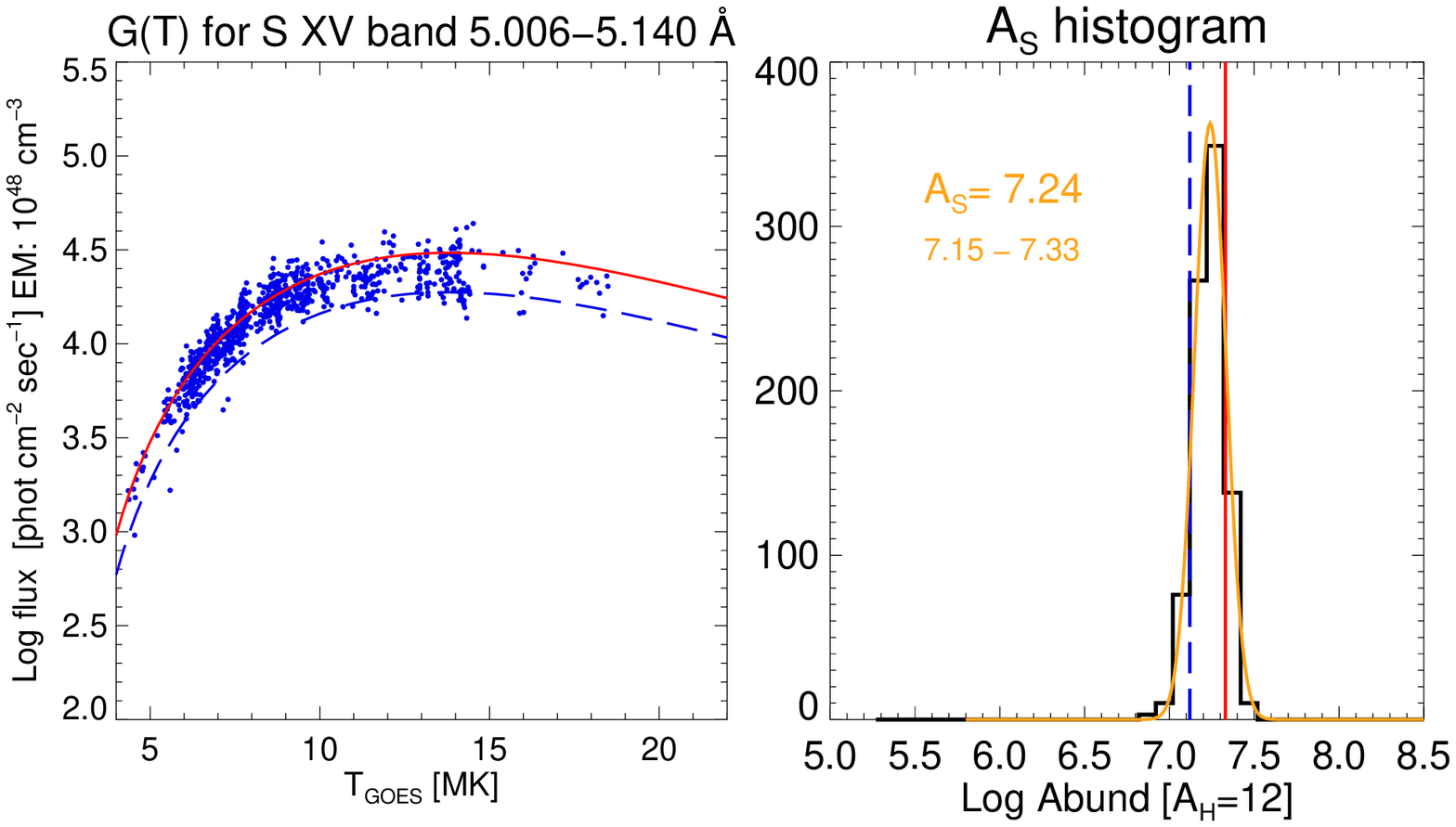}
\caption{Left panel: Measured \ion{S}{15} line emission (RESIK channel~4 range 5.006--5.140~\AA) divided by EM$_{GOES}$ plotted against $T_{\rm GOES}$. Right panel: Histogram of abundance determinations with best-fit Gaussian curve based on the blue dots. The value of $A({\rm S}) = 7.24 \pm 0.08$.  } \label{SXV_w_GofT_ch4}
\end{figure}

The \ion{S}{15} $1s^2 - 1s2s$, $1s^2 - 1s2p$ triplet of lines near 5~\AA, falling at the short-wavelength end of channel~4, are among the most intense of all lines in RESIK spectra, but because they fall in the least well characterized channel, abundance estimates may be slightly compromised. As already indicated, the resonance ($w$) line is unresolved from the intercombination ($x$, $y$) lines and dielectronic lines are also unresolved. The total line flux in the 1448 spectra available from the thirteen flares given in Table~\ref{flare_list} was estimated by taking the flux in the range 5.006--5.140~\AA\ and subtracting the background in the 5.000--5.005~\AA\ and 5.248--5.548~\AA\ intervals on either side of the line emission. The estimated sulphur abundance, $A({\rm S}) = 7.24 \pm 0.09$, is greater than those from the \ion{S}{15} $w3$ and $w4$ lines. This may be due to the fact that the \ion{S}{15} triplet falls close to the edge of the channel~4 crystal, where the crystal radius of curvature may be larger and the resulting increased crystal reflectivity leads to a slight increase in the fluxes of the lines. Also, as the channel~4 range is more crowded with intense spectral lines (mostly due to Si at wavelengths greater than 5.15~\AA) and the dispersion is lower owing to the smaller radius of crystal curvature, a line-free portion of background is more difficult to define than for the \ion{S}{15} $w3$ and $w4$ lines, and so there may be additional slight uncertainty in the line flux estimates. This is not, however, particularly evident from the plots of Figure~\ref{SXV_w_GofT_ch4}.

% Fig. 5: S XV triplet lines, channel 4

\begin{figure}
\epsscale{.60}
\plotone{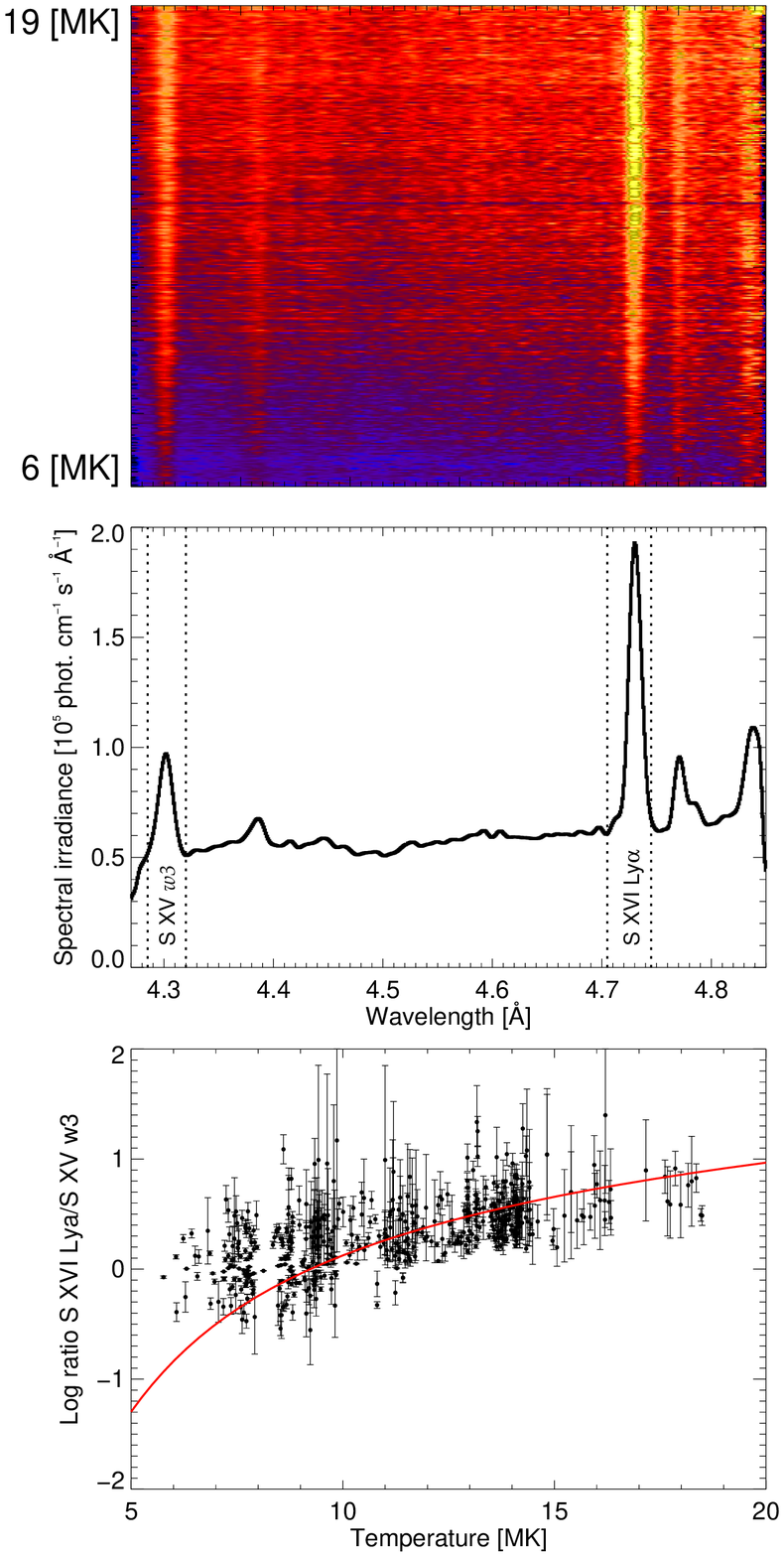}
\caption{Top panel: The 725 RESIK channel~3 spectra in eight flares for which the \ion{S}{15} $w3$ line fell in this channel at its low-wavelength end, stacked in order of $T_{GOES}$. Center panel:  Average of these spectra. Bottom panel: Measured flux ratio of \ion{S}{16} Ly-$\alpha$ to \ion{S}{15} $w3$ lines with statistical uncertainties (dots with error bars) plotted against $T_{GOES}$ and theoretical variation with temperature from {\sc chianti}. } \label{SXVI_Lya_SXV_w3_ratio_flares}
\end{figure}

\begin{figure}
\epsscale{.80}
\plotone{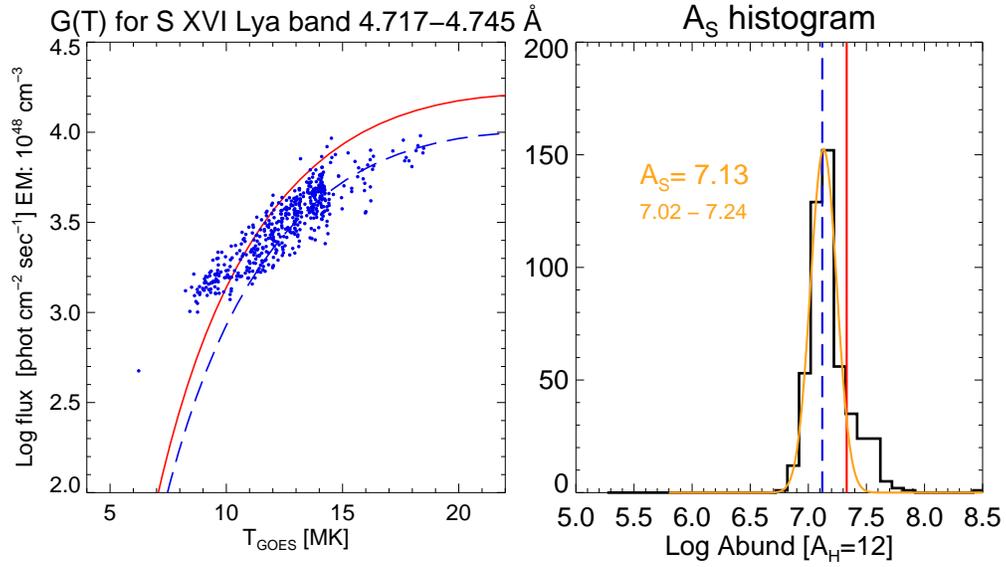}
\caption{Left panel: Measured \ion{S}{16} Ly$\alpha$ line emission (RESIK channel~3 range 4.717--4.745~\AA) divided by EM$_{GOES}$ plotted against $T_{\rm GOES}$. As \ion{S}{16} Ly-$\alpha$ line is weak for low temperatures, only spectra with total photon counts in the line $> 100$ are selected. Right panel: Histogram of S abundance determinations with  best-fit Gaussian curve. Here the value of $A({\rm S}) = 7.13 \pm 0.09$.  } \label{SXVI_GofT_ch3}
\end{figure}

Figure~\ref{SXVI_Lya_SXV_w3_ratio_flares} (top panel) shows the 725 spectra for the eight flares for which $w3$ fell in channel~3 stacked in order of the value of $T_{GOES}$ (increasing upwards). The center panel shows the averaged spectrum. Both the $w3$ and \ion{S}{16} Ly-$\alpha$ lines are measurable over the full temperature range. In Figure~\ref{SXVI_Lya_SXV_w3_ratio_flares} (bottom panel), the logarithm of the flux ratios \ion{S}{16} Ly-$\alpha$ to \ion{S}{15} $w3$ are plotted against $T_{GOES}$ with statistical uncertainties. The solid curve is the theoretical ratio calculated from {\sc chianti} using contribution functions that include the most significant of unresolved dielectronic satellite lines. There is approximate agreement, to within the uncertainties, for $T_{GOES} \gtrsim 10$~MK. For smaller temperatures the points increasingly depart from the theoretical curve. At present, this departure is unexplained but we are considering different physical mechanisms that may be responsible for this observed discrepancy, such as uncertainties in the ionization fractions of H-like and He-like S ions or the presence of high-energy non-Maxwellian tails in plasma where the X-ray lines are formed. Sulphur abundance estimates can be made from the \ion{S}{16} Ly-$\alpha$ line emission alone; this is indicated by Figure~\ref{SXVI_GofT_ch3} where, following other plots, we show the line flux divided by $EM_{GOES}$ plotted against $T_{GOES}$. This is done for total photon counts in the \ion{S}{16} Ly-$\alpha$ line exceeding 100. The tendency for the points to lie above the theoretical $G(T_e)$ curve at low temperatures is again evident. The peak of the abundance distribution (Figure~\ref{SXVI_GofT_ch3}, right panel) indicates a value for $A({\rm S}) = 7.13 \pm 0.11$, very similar to the estimates from the \ion{S}{15} $w3$ and $w4$ lines.

% Fig. 6: S XVI Ly-alpha to S XV w3 line ratio + stacked spectra for 8 flares (725 spectra)

% Fig. 7: S XVI Ly-alpha, channel 3, 725 spectra from 8 flares.

% Fig. 8: S XV w4 / Ar XVII w RESIK spectra, channel 2
\begin{figure}
\epsscale{.80}
\plotone{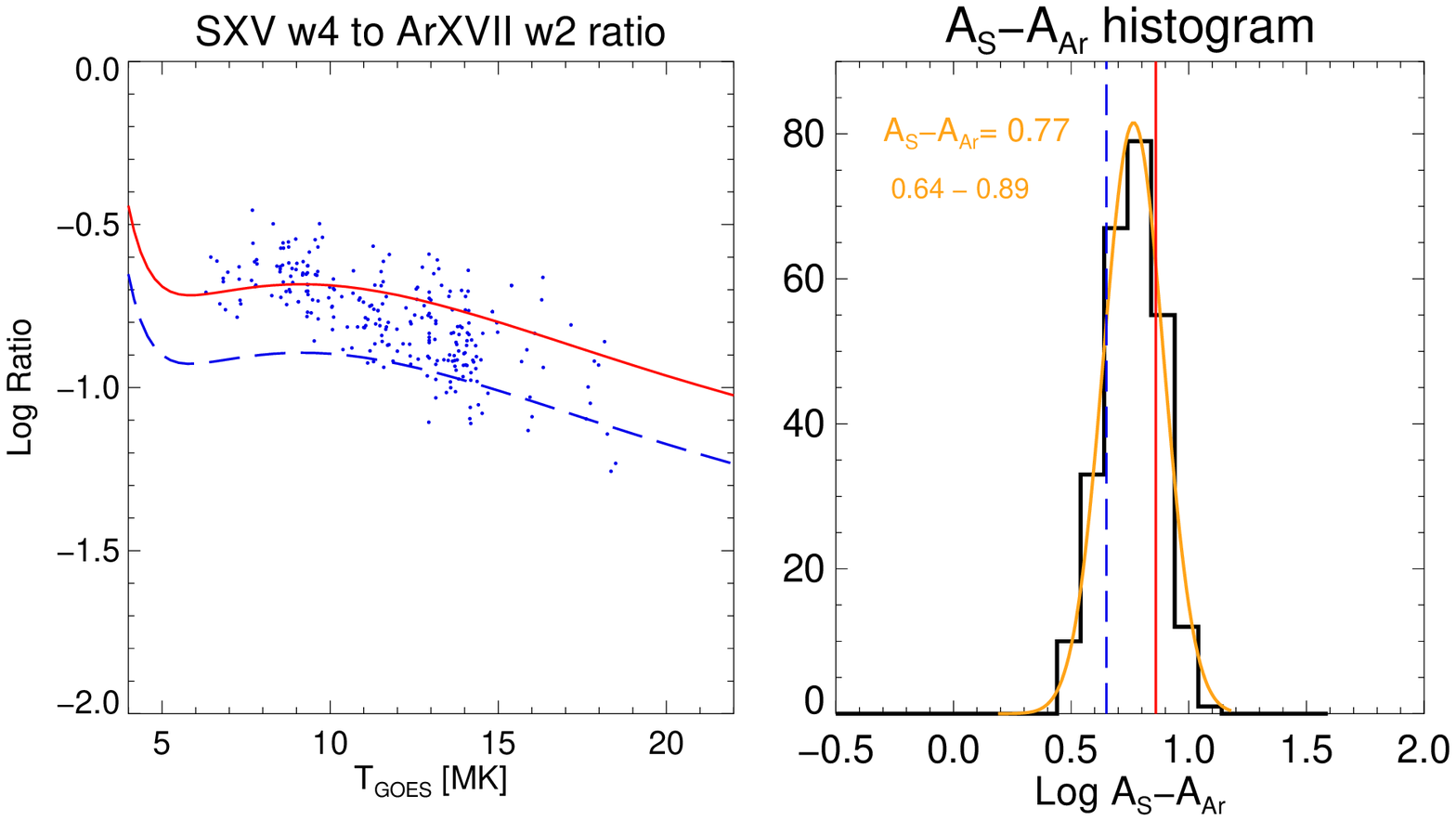}
\caption{Left panel: Ratio of measured \ion{S}{15} $w4$ line emission (4.075--4.095~\AA) to \ion{Ar}{17} line emission (3.94--4.01~\AA) in RESIK channel~2. Spectra have been selected such that the total counts in the \ion{S}{15} $w4$ line are $> 50$ and the \ion{Ar}{17} $w$ line are $> 200$. Right panel: Histogram of determinations of $A({\rm S}) - A({\rm Ar})$; the best-fit Gaussian indicates that the ratio is $A({\rm S}) - A({\rm Ar}) = + 0.77 \pm 0.11$. With $A({\rm Ar}) = 6.46 \pm 0.08$, $A({\rm S})$ is estimated to be $7.23 \pm 0.14$.} \label{SXV_w4_ArXVII_w_ratio}
\end{figure}

Another means of determining $A({\rm S})$ is offered by the line ratio of the \ion{S}{15} $w4$ line to the nearby \ion{Ar}{17} $1s^2 - 1s2l$ lines (in the 3.94--4.01~\AA\ range), both in channel~2. The Ar abundance in RESIK flares was discussed in previous work \citep{syl10c}. An improved analysis procedure has resulted in an estimated Ar abundance that is slightly revised upward from this work, $A({\rm Ar}) = 6.46 \pm 0.09$ (previously $A({\rm Ar}) = 6.44 \pm 0.07$). The measured ratios of the \ion{S}{15} $w4$ to the \ion{Ar}{17} $w$ lines are plotted against $T_{GOES}$ in Figure~\ref{SXV_w4_ArXVII_w_ratio}, together with theoretical curves on the assumption of a photospheric and coronal S abundance. The result of this analysis is $A({\rm S}) - A({\rm Ar}) = + 0.77 \pm 0.12$. With the presently determined Ar abundance, this gives $A({\rm S}) = 7.23 \pm 0.14$.

% Section 4
\section{SUMMARY AND CONCLUSIONS}

This analysis of highly ionized sulphur lines appearing in RESIK flare spectra has led to several estimates of the sulphur abundance which can be compared with each other and with previous values. We consider the estimate from the \ion{S}{15} $w4$ line, viz. $A({\rm S}) = 7.16 \pm 0.17$, to be the most reliable as the line occurs in RESIK's channel~2 which is better characterized than channels~3 and 4 where the other sulphur lines appear in that solar continuum practically uncontaminated by crystal fluorescence can be measured. By measuring the \ion{S}{15} $w4/w3$ line flux ratio and comparing its temperature variation with theory (this is independent of S abundance and ionization equilibrium calculations), we have established that though the $w3$ line only ever appears in channel~3 at its short-wavelength edge possible anomalous reflectivities due to crystal curvature effects can be neglected. This then enables another estimate of the sulphur abundance, $A({\rm S}) = 7.17 \pm 0.20$, similar as expected to that from the $w4$ line. The \ion{S}{15} $1s^2 - 1s2s$, $1s^2 - 1s2p$ line triplet near 5~\AA\ (together with unresolved \ion{S}{14} dielectronic satellites), occurring in channel~4, leads to the estimated sulphur abundance $A({\rm S}) = 7.24 \pm 0.09$. The small statistical uncertainty in this estimate does not necessarily indicate a high precision, since channel~4 may have a somewhat greater uncertainty in its relative calibration and the subtraction of background is subject to more uncertainty in this case. The \ion{S}{16} Ly-$\alpha$ line in channel~3 gives $A({\rm S}) = 7.13 \pm 0.11$, though there is some uncertainty in this through observations at lower temperatures. Finally, a sulphur-to-argon abundance ratio is provided by the \ion{S}{15} $w4$ to \ion{Ar}{17} $1s^2 - 1s2s$, $1s^2 - 1s2p$ line group; with an Ar abundance, $A({\rm Ar}) = 6.46 \pm 0.09$, slightly revised from previous work \citep{syl10c}, this leads to $A({\rm S}) = 7.23 \pm 0.14$.

The estimate from the \ion{S}{15} $w4$ line, $A({\rm S}) = 7.16 \pm 0.17$, is considered by us to be the most reliable. It is very similar to the estimates from the \ion{S}{15} $w3$ and \ion{S}{16} Ly-$\alpha$ lines, and while less than that from the \ion{S}{15} 5~\AA\ lines, the estimate from the latter may be subject to more uncertainty through background subtraction and the fact that channel~4 is less well characterized than channels~2 and 3. It is less by a factor 1.5 than the coronal value of \cite{fel00}, but is in very close agreement with the photospheric values of \cite{caf11} (7.16) and \cite{asp09} (7.12) as well as the quiet-time solar wind measurements (7.15: \cite{rea99}) and meteoritic value (7.19: \cite{lod03}). On a FIP-dependent model for abundance enhancements over photospheric abundances, this argues strongly for sulphur, despite its ``intermediate" value of FIP (10.36~eV) being a high-FIP element, similar to Ar, the abundance of which we also estimated from RESIK spectra.

We acknowledge financial support from the Polish Ministry of Education and Science Grant 2011/01/B/ST9/05861 and the UK Royal Society/Polish Academy of Sciences International Joint Project for travel support.  The research leading to these results has received funding from the European Commission's Seventh Framework Programme (FP7/2007-2013) under the grant agreement eHEROES (project number 284461). {\sc chianti} is a collaborative project involving Naval Research Laboratory (USA), the Universities of Florence (Italy) and Cambridge (UK), and George Mason University (USA).


\begin{thebibliography}{}

\bibitem[Aggarwal \& Kingston(1991)]{agg91} Aggarwal, K. M., \& Kingston, A. E. 1991, J. Phys. B, 24, 4583

\bibitem[Asplund et al.(2009)]{asp09} Asplund, M., Grevesse, N., Sauval, A. J., \& Scott, P. 2009, \araa, 47, 481

\bibitem[Bryans et al.(2009)]{bry09} Bryans, P., Landi, E., \& Savin, D. W. 2009, \apj, 691, 1540

%\bibitem[Caffau et al.(2007)]{caf07} Caffau, E., Faraggiana, R., Bonifacio, P., Ludwig, H.-G., Steffen, M. 2007a, \aap, 470, 699

\bibitem[Caffau et al.(2011)]{caf11} Caffau, E., Ludwig, H.-G., Steffen, M., Freytag, B., \& Bonifacio, P. 2011, \solphys, 268, 255

\bibitem[Dere et al.(1997)]{der97} Dere, K. P., Landi, E., Mason, H. E., Monsignori Fossi, B. C., \& Young, P. R. 1997, \aaps, 125, 149

\bibitem[Dere et al.(2009)]{der09} Dere, K. P., Landi, E., Young, P. R., Del Zanna, G., Landini, M., \& Mason, H. E. 2009, \aap, 498, 915

\bibitem[Feldman(1992)]{fel92} Feldman, U. 1992, Phys. Scr., 46, 202

\bibitem[Feldman \& Laming(2000)]{fel00} Feldman, U., \& Laming, J. M. 2000, Phys. Scr., 61, 222

\bibitem[Fludra \& Schmelz(1999)]{flu99} Fludra, A, \& Schmelz, J. T. 1999, \aap, 348, 286

\bibitem[Grevesse \& Sauval(1998)]{gre98} Grevesse, N., \& Sauval, A. J. 1998, \ssr, 85, 161

\bibitem[Harra-Murnion et al.(1996)]{har96} Harra-Murnion, L. K., et al. 1996, \aap, 306, 670

\bibitem[H\'enoux(1998)]{hen98} H\'enoux, J.-C. 1998, \ssr, 85, 215

\bibitem[Kimura et al.(2000)]{kim00} Kimura, E., Nakazaki, S., Berrington, K. A., \& Norrington, P. H. 2000, J Phys B, 33, 3449

\bibitem[Laming(2004)]{lam04} Laming, J. M. 2004, \apj, 614, 1063

\bibitem[Laming(2009)]{lam09} Laming, J. M. 2009, \apj, 695, 954

\bibitem[Laming(2012)]{lam12} Laming, J. M. 2012, \apj, 744, 115

\bibitem[Lodders(2003)]{lod03} Lodders, K. 2003, \apj, 591, 1220

\bibitem[Phillips \& Dennis(2012)]{phi12} Phillips, K. J. H., \& Dennis, B. R. 2012, \apj, 748, 52

\bibitem[Phillips et al.(2003)]{phi03} Phillips, K. J. H., Sylwester, J., Sylwester, B., \& Landi, E. 2003, \apj, 589, L113

\bibitem[Phillips et al.(2008)]{phi08} Phillips, K. J. H., Feldman, U., \& Landi, E. 2008, Ultraviolet and X-ray Spectroscopy of the Solar Atmosphere (chap. 4, Cambridge: Cambridge Univ. Press)

\bibitem[Phillips et al.(2010)]{phi10} Phillips, K. J. H., Sylwester, J., Sylwester, B., \& Kuznetsov, V. D. 2010, \apj, 711, 179 % continuum paper

\bibitem[Reames(1999)]{rea99} Reames, D. V. 1999, \apj, 518, 473

\bibitem[Sylwester et al.(2005)]{syl05} Sylwester, J., Gaicki, I., Kordylewski, Z., et al. 2005, \solphys, 226, 45

\bibitem[Sylwester et al.(2011)]{syl11} Sylwester, B., Phillips, K. J. H., Sylwester, J., \& Kuznetsov, V. D. 2011, \apj, 738, 49 % Cl abund.

\bibitem[Sylwester et al.(2010a)]{syl10a} Sylwester, B., Sylwester, J., \& Phillips, K. J. H. 2010a, \aap, 514, 82 % AR abunds.

\bibitem[Sylwester et al.(2010b)]{syl10b} Sylwester, J., Sylwester, B., Phillips, K. J. H., \& Kuznetsov, V. D. 2010b, \apj, 710, 804 % K abund.

\bibitem[Sylwester et al.(2010c)]{syl10c} Sylwester, J., Sylwester, B., Phillips, K. J. H., \& Kuznetsov, V. D. 2010c, \apj, 720, 1721 % Ar abund.

\bibitem[Von Steiger et al.(2000)]{von00} Von Steiger, R. et al. 2000, \jgr, 105, 27,217

\bibitem[Watanabe et al.(1995)]{wat95} Watanabe, T., et al. 1995, \solphys, 157, 169

\end{thebibliography}
\end{document}